\documentclass[11pt,letterpaper,twoside,notitlepage,twocolumn]{amsart}

\usepackage{amsmath}%
\usepackage{amsfonts}%
\usepackage{amssymb}%
\usepackage{graphicx}
\usepackage[margin=10pt,font={small,it},labelfont=bf]{caption}
\usepackage{fancyhdr}
\usepackage{fancybox}
\usepackage{fancybox}
\usepackage{tabls}
\usepackage[left=2.75cm,top=2cm,right=2cm,bottom=2cm,includehead]{geometry}
\usepackage{cite}

\newtheorem*{thm}{Theorem}

\numberwithin{equation}{section}

\newcommand{\equ}[1]{\begin{equation} #1 \end{equation}} 
\newcommand{\eqa}[1]{\begin{eqnarray} #1 \end{eqnarray}} 
\newcommand{\ci}{\mathrm i} 
\newcommand{\td}{\mathrm d} 
\newcommand{\mr}{\mathrm} 
\newcommand{\rb}[1]{\left( #1 \right)} 
\newcommand{\cb}[1]{\left\{ #1 \right\}}
\newcommand{\ang}[1]{\left\langle #1 \right\rangle} 
\newcommand{\stb}[1]{\left[ #1 \right]}

\newcommand{\nn}{\nonumber} 
\newcommand{\der}[1]{\frac{\partial}{\partial #1}}

\begin{document}

\title{On invariants and scalar chiral correlation functions in $\mathcal{N}=1$ superconformal field theories\vspace{-0.3cm}}
\author{Holger Knuth\vspace{-0.3cm}}
\twocolumn[
  \begin{@twocolumnfalse}
\begin{flushright}
      \texttt{arXiv:1010.2740v2}\\
      \texttt{LQP:10101400}\\
\end{flushright}
\begin{abstract}
A general expression for the four-point function with vanishing total R-charge of anti-chiral and chiral superfields in $\mathcal{N}=1$ superconformal theories is given. It is obtained by applying the exponential of a simple universal nilpotent differential operator to an arbitrary function of two cross ratios. To achieve this the nilpotent superconformal invariants according to Park are focused. Several dependencies between these invariants are presented, so that eight nilpotent invariants and $27$  monomials of these invariants of degree $d > 1$ are left being linearly independent. It is analyzed, how terms within the four-point function of general scalar superfields cancel in order to fulfill the chiral restrictions. 
\end{abstract}
\maketitle
\vspace{-1.2cm}
\end{@twocolumnfalse}
]

\let\thefootnote\relax\footnotetext{\emph{Key words and phrases}: Superconformal symmetry; Correlation functions; Chiral superfields.\\ PACS 2010: 11.30.Pb; 11.25.Hf\\MSC 2010: 81T60.}

\markboth{{\sc Holger Knuth}}{{\sc On invariants and scalar chiral correlation functions in $\mathcal{N}=1$ SCFT}}

\sloppy
\section{Introduction}
In the early beginnings of supersymmetry \cite{Wess:1974tw,springerlink:10.1007/BF01079417,springerlink:10.1007/BF01028617,springerlink:10.1007/BF01036669,springerlink:10.1007/BF01039108} four-dimensional $\mathcal{N}=1$ superconformal field theories came into the focus of several researchers. But due to the mainly perturbative approach to supersymmetric field theories, in which non-trivial theories could not sustain conformal invariance, the mainstream attached to super-Poincar\'{e} symmetry. 

Still insights in the conformal invariance of supersymmetric non-Abelian gauge theories at renormalization group fixed points with vanishing $\beta$-function (\cite{Seiberg:1994pq} and more recently \cite{Intriligator:2003jj,Intriligator:2003mi,Barnes:2005zn}) brought new interest to a non-perturbative analysis of superconformal field theories in the mid-90s. The very recent study of new supergravity models with the potential to resolve some problems of minimal supersymmetric standard models \cite{Ferrara:2010yw,Ferrara:2010in} could possibly again renew attention on this topic. Here big emphasis is given to the underlying superconformal symmetry.

Chiral superfields gathered most attention not only because of their prominent role in supersymmetry through the construction of F-Term Lagrangians and as suitable multiplets containing elementary particles, but also for the sake of their relative simplicity. This paper focuses on chiral and anti-chiral superfields as well, as it comes to correlation functions in the second half. 

In the eighties the two authors of \cite{Dobrev:1985qv,Dobrev:1985vh,Dobrev:1985qz} concentrated on a group theoretical approach to extended conformal supersymmetry. 

A later series of papers started also with results for chiral superfield in $\mathcal{N}=1$ \cite{Howe:1995aq,Howe:1996pw,Howe:1996rb}, but it focused afterwards the analysis of theories with extended superconformal symmetry on analytic superspace (e.g. \cite{Howe:1996rm,Howe:1998zi,Eden:2000gg,Eden:2000qp,Heslop:2002hp,Heslop:2003xu}). This approach is purely on-shell. It was shown, that there are no nilpotent invariants of up to four points of analytic superspace\cite{Eden:1999gh,Howe:1999hz}. 

Another line of work concentrated first on correlation functions of only chiral superfields for  $\mathcal{N}=1$\cite{Osborn:1998qu,Dolan:2000uw}. While the two point function is a pure contact term, the three point function is only consistent with the Ward identities, if the total R-charge is one. Later papers then mainly dealt with $\mathcal{N}=4$  \cite{Arutyunov:2001mh,Dolan:2001tt,Dolan:2002zh,Dolan:2003hv,Dolan:2004iy}. 

In a more general approach implications of superconformal symmetry on correlation functions of arbitrary quasi-primary fields were derived by Park \cite{Park:1997bq,Park:1999pd}. For that purpose all invariants of the $\mathcal{N}$-extended superconformal group for all $\mathcal{N}\geq 1$ were constructed on ordinary super Minkowski space. It is stated, that correlation functions with vanishing R-charge are functions of these invariants. For the case of non-vanishing R-charge the correlation functions are nilpotent and depend on a larger set of invariants, which are not required to be invariant under the $U(1)_Y$ symmetry. Here only the former case is discussed.

Section \ref{susp} takes a look at the coordinates of superspace and variables, which are suitable for the construction of invariants and expression of n-point functions because of their homogeneous transformation properties. 

In section \ref{inv} the three- and four-point invariants for $\mathcal{N}=1$ from \cite{Park:1999pd} are given. It is argued, that there are only 10 independent invariants of four points: two cross ratios and eight nilpotent invariants. This part is central to this paper because it shows, that in $\mathcal{N}=1$ supersymmetry fewer invariants are needed to express any superconformal four-point function than thought hitherto (\cite{Park:1999pd} argues it to be maximally $16$).

Section \ref{sufi} shortly introduces scalar chiral and anti-chiral superfields, before their correlation functions with vanishing R-charge are discussed in section \ref{corr}. While two- and three-point functions do not need a long discussion, the discussion of four-point functions is arranged around the proof of a theorem giving their general form. 

The theorem gives the four-point function as an arbitrary function of two superconformal cross ratios containing all model specific information, to which a universal differential operator is applied. Its proof starts with the general form of scalar four-point functions from \cite{Park:1999pd} and so gives an insight in  the elimination of terms needed, so that the four-point function only depends on the correct set of chiral- and anti-chiral coordinates. While it was already known, that this function can be written in terms of two non-nilpotent invariants, this theorem answers the question, how these chiral scalar four-point functions are expressed in terms of the set of invariants chosen in section \ref{inv}. 

At last a short calculation derives another way to write rational superconformal four-point functions as a power series. This allows to directly transfer results from global conformal field theory to global superconformal field theory, which is the reason, why we prefer this expression to those known so far.

\section{Superspace}
\label{susp} Supersymmetry is generated by operators $Q_{\alpha}$ and their hermitian conjugates $Q_{\alpha}^{\dagger}=\bar{Q}_{\dot{\alpha}}$. They are two-component spinors like the fermionic component fields, as they relate these fields to bosonic fields and vice versa. 
Together with $P_{\mu}=\rb{H, -\mathbf{P}}$ and $M_{\mu\nu}$ they fulfill the four-dimensional $\mathcal{N}=1$ supersymmetry algebra. The only non-vanish\-ing (anti-)commutators are 
\eqa{
\cb{Q_{\alpha},\bar{Q}_{\dot{\alpha}}}&=&2\sigma_{\alpha\dot{\alpha}}^{\mu}P_{\mu}
}
and the commutation relations containing $M_{\mu\nu}$, which express the spinoral, vectorial and tensorial transformation properties of the generators. With this the general element of the translational supergroup is given by
\equ{
g(z)=e^{\ci\rb{x^{\mu}P_{\mu}+\theta^{\alpha}Q_{\alpha}+\bar{Q}_{\dot{\alpha}}\bar{\theta}^{\dot{\alpha}}}}\; ,
}
where the parameters $z^M=\rb{x^{\mu},\theta^{\alpha},\bar{\theta}^{\dot{\alpha}}}$ are the coordinates of superspace. 
$\theta^{\alpha}$ and $\bar{\theta}^{\dot{\alpha}}$ are Gra\ss mann-valued spinors. The actual reason for the results in this paper to be only valid for $\mathcal{N}=1$ is the property
\eqa{\label{ttequ1}
\theta^{\alpha}\theta^{\beta}&=&-\frac{1}{2}
\epsilon^{\alpha\beta}\theta\tilde{\theta}\;,\\ 
\label{ttequ2} \bar{\theta}^{\dot{\alpha}}\bar{\theta}^{\dot{\beta}}&=&
\frac{1}{2}
\bar{\epsilon}^{\dot{\alpha}\dot{\beta}}
\tilde{\bar{\theta}}\bar{\theta}
}
with the anti-symmetric $2\times 2$ matrices $\epsilon^{\alpha\beta}$ and $\bar{\epsilon}^{\dot{\alpha}\dot{\beta}}$ with $\epsilon_{12}=\epsilon^{21}=\bar{\epsilon}_{\dot{1}\dot{2}}=\bar{\epsilon}^{\dot{2}\dot{1}}=1$. The exterior derivative on superspace is given by
\eqa{
\td&=&\rb{\td x^{\mu}+\ci \td\theta \sigma^{\mu} \bar{\theta} -\ci \theta \sigma^{\mu} \td\bar{\theta}}\der{x^{\mu}}\\
\nn &&+\td\theta^{\alpha}D_{\alpha}-\td\bar{\theta}^{\dot{\alpha}}\bar{D}_{\dot{\alpha}}
} 
with the covariant derivatives$^1$ \footnote{$^1$The spinoral indices will be omitted in most of the paper and a tilde is used to indicate lower indices of the spinors.} 
\eqa{
D&=&\der{\theta}- \ci \sigma^{\mu}\bar{\theta} \der{x^{\mu}}\; ,\\
\bar{D}&=&-\der{\bar{\theta}}+ \ci \theta \sigma^{\mu}\der{x^{\mu}}\; .
}
For the discussion of chiral and anti-chiral superfields it is convenient to use chiral and anti-chiral coordinates, which are respectively defined by
\eqa{
{x_+}^{\mu}&=&x^{\mu}-\ci \theta \sigma^{\mu} \bar{\theta}\; ,\\
{x_-}^{\mu}&=&x^{\mu}+\ci \theta \sigma^{\mu} \bar{\theta}\; .
}
The covariant derivatives for these coordinates are
\eqa{
\label{Dplus}\quad \quad \bar{D}_+=-\der{\bar{\theta}} \; ,& D_+=\der{\theta}-2 \ci \sigma^{\mu}\bar{\theta} \der{{x^-}^{\mu}}\; ,\\
\label{Dminus}D_-=\der{\theta} \; ,&  \bar{D}_-=-\der{\bar{\theta}}+2 \ci \theta \sigma^{\mu}\der{{x^+}^{\mu}}\; .
}
As we will see in the following sections the correlation functions are expressed in terms of the following intervals
\eqa{
\label{xibj} x_{\bar{i}j}^{\mu}&=&{x_{i-}}^{\mu}-{x_{j+}}^{\mu}-2 \ci \theta_j \sigma^{\mu} \bar{\theta_i}\; ,\\
\theta_{ij}^{\alpha}&=&\theta_i^{\alpha} - \theta_j^{\alpha}\; ,\\
\bar{\theta}_{ij}^{\dot{\alpha}}&=&\bar{\theta}_i^{\dot{\alpha}}-\bar{\theta}_j^{\dot{\alpha}}\; ,
}
which transform homogeneously under superconformal transformations (cf. \cite{Park:1999pd}). Writing the interval \eqref{xibj} as hermitian $2\times 2$ matrices by contraction with Pauli matrices, 
${\mathrm x_{\bar{i}j}}_{\alpha\dot{\alpha}}=x_{\bar{i}j}^{\mu}{\sigma_{\mu}}_{\alpha\dot{\alpha}}$ 
and ${\tilde{\mathrm x}_{\bar{i}j}}^{\dot{\alpha}\alpha}=x_{\bar{i}j}^{\mu}{\tilde{\sigma}_{\mu}}^{\dot{\alpha}\alpha}$, their inverses are
\eqa{
{\mathrm x_{\bar{i}j}}^{-1}= \frac{\tilde{\mathrm x}_{\bar{i}j}}{{x_{\bar{i}j}}^2}\; ,&&
{\tilde{\mathrm x}_{\bar{i}j}^{-1}}= \frac{\mathrm x_{\bar{i}j}}{{x_{\bar{i}j}}^2}\; .
}

\section{Superconformal invariants}
\label{inv}  In preparation of the discussion of the correlation functions the invariants of up to four points in superspace have to be constructed. This was done together with a detailed analysis of superconformal symmetry and correlation functions in \cite{Park:1999pd}. Here the results are summarized and relations are found between invariants of four points. These reduce the supposed number of independent invariants, which have to be considered in the expression of the four-point function, from $16$ to $10$. Eight of them are nilpotent and form $36$ linear independent monomials including those of degree $0$ and $1$.
\subsection{Invariant of three points}
\label{sec3pI}
There are no invariants of two points, but one of three points. This three point invariant can be expressed in terms of
\eqa{
\label{X1plus}\tilde{\mr X}_{1+}&=&-{\mr x_{\bar{1}3}}^{-1} \mr x_{\bar{2}3} {\mr x_{\bar{2}1}}^{-1}\; , \\
\label{X1minus}\tilde{\mr X}_{1-}&=&{\mr x_{\bar{1}2}}^{-1} \mr x_{\bar{3}2} {\mr x_{\bar{3}1}}^{-1}\; .
%\mr X_{1+}&=&{\mr x_{\bar{2}1}}^{-1} \mr x_{\bar{2}3} {\mr x_{\bar{1}3}}^{-1}\; , \\
%\mr X_{1-}&=&-{\mr x_{\bar{3}1}}^{-1} \mr x_{\bar{3}2} {\mr x_{\bar{1}2}}^{-1}
}
The squares of the four-vectors corresponding to $\tilde{\mr X}_{1+}$ and $\tilde{\mr X}_{1-}$ are needed:
\eqa{
\label{X1pm} X_{1+}^2=\frac{x_{\bar{2}3}^2}{x_{\bar{2}1}^2x_{\bar{1}3}^2}\; , &&  
X_{1-}^2=\frac{x_{\bar{3}2}^2}{x_{\bar{3}1}^2x_{\bar{1}2}^2}\; .
}
With these the three-point invariant is
\eqa{\label{I3PFJ1}
I_{3PF}&=&J_1-1 \\
\label{J1} \text{with}\quad J_1&=&\frac{X_{1+}^2}{X_{1-}^2} \; .
}
$I_{3PF}$ is nilpotent because the squares of the intervals, $x_{\bar{i}j}^2$, become symmetric in their indices, if the Gra\ss mann variables are set to zero, and so $J_1$ is one in this case. \\
Another possibility to express this invariant is the use of two Gra\ss mann spinors,
\eqa{
\Theta_{1}&=&\ci\rb{\tilde{\bar{\theta}}_{21}{{\mr x}_{\bar{2}1}}^{-1}-\tilde{\bar{\theta}}_{31}{{\mr x}_{\bar{3}1}}^{-1}}\; ,\\
\bar{\Theta}_{1}&=&\ci\rb{{{\mr x}_{\bar{1}3}}^{-1}\tilde{\theta}_{13}-{{\mr x}_{\bar{1}2}}^{-1}\tilde{\theta}_{12}}\; ,
}
and either $\tilde{\mr X}_{1+}$ or $\tilde{\mr X}_{1-}$, because
\equ{\label{X1plusX1min}
\tilde{\mr X}_{1-}=\tilde{\mr X}_{1+}-4\ci \bar{\Theta}_{1}\Theta_{1}\; .
}
With the help of this equation one has
\equ{\label{I3pAlt}
I_{3PF}=-4\ci \frac{{\Theta}_{1}{\mr X}_{1+}\bar{\Theta}_{1}}{ X_{1+}^2}\; .
}

\subsection{Invariants of four points}
\label{inv4p}
The construction of four-point invariants leads to both non-nilpotent and nilpotent invariants.  
\subsubsection{Cross ratios}
The cross ratios, 
\equ{\label{croratgen}
\frac{x_{\bar{r}s}^2 x_{\bar{t}u}^2}{x_{\bar{r}u}^2 x_{\bar{t}s}^2}\; ,
}
reduce for vanishing Gra\ss mann variables to the ordinary conformal cross ratios. 

Defining analogously to \eqref{X1pm}
\eqa{
\label{X1ipm}\quad X_{1(i-1)+}^2=\frac{x_{\bar{i}4}^2}{x_{\bar{i}1}^2x_{\bar{1}4}^2}\; , \!\!\!&&\!\!\! X_{1(i-1)-}^2=\frac{x_{\bar{4}i}^2}{x_{\bar{4}1}^2x_{\bar{1}i}^2}
}
with $i=1,2$, one gets the nilpotent invariant of three points -- ignoring the added constant $1$ -- with the corresponding indices replaced: 
\equ{\label{J112}
J_{1(i)}=\frac{X_{1(i)+}^2}{X_{1(i)-}^2}\; .
}
They relate the cross ratios with each other. There are six different cross ratios and their inverses (cf. eq. \eqref{croratgen}).
 
Picking out the two cross ratios,
\equ{\label{crorat}
\mathcal{I}_1=\frac{x_{\bar{1}4}^2 x_{\bar{3}2}^2}{x_{\bar{1}2}^2 x_{\bar{3}4}^2}\; ,
 \quad \mathcal{I}_2=\frac{x_{\bar{1}4}^2 x_{\bar{2}3}^2}{x_{\bar{1}3}^2 x_{\bar{2}4}^2}\; ,
}
the other cross ratios are
\eqa{\label{croratfur1}
\mathcal{I}_3&=\frac{x_{\bar{3}2}^2 x_{\bar{4}1}^2}{x_{\bar{3}1}^2 x_{\bar{4}2}^2}=&J_1^{-1} J_{1(1)} \mathcal{I}_2\;, \\
\mathcal{I}_4&=\frac{x_{\bar{2}4}^2 x_{\bar{3}1}^2}{x_{\bar{2}1}^2 x_{\bar{3}4}^2}=&J_1 \mathcal{I}_1 \mathcal{I}_2^{-1}\;, \\
\mathcal{I}_5&=\frac{x_{\bar{2}3}^2 x_{\bar{4}1}^2}{x_{\bar{2}1}^2 x_{\bar{4}3}^2}=&J_1 \mathcal{I}_1 \mathcal{I}_2^{-1}\;, \\
\label{croratfur4}\mathcal{I}_6&=\frac{x_{\bar{1}3}^2 x_{\bar{4}2}^2}{x_{\bar{1}2}^2 x_{\bar{4}3}^2}=&J_1 J_{1(1)}^{-1} J_{1(2)} \mathcal{I}_1 \mathcal{I}_2^{-1}\;.
}

There are no other non-nilpotent invariants, which cannot be expressed by $\mathcal{I}_1$, $\mathcal{I}_2$ and nilpotent invariants, because the number of independent non-nilpotent invariants has to be the same as the number of ordinary conformal invariants.
\subsubsection{Nilpotent invariants}
\label{nilpot4pi} 
There are many possible sets of eight independent nilpotent invariants one can choose. Here the choice is 
\eqa{
\label{nilpotinv1}I_{1ij}&=&-\frac{\tilde{\bar{\Theta}}_{1(j)}\tilde{\mr X}_{1(1)+}\tilde{\Theta}_{1(i)}}{\sqrt{X_{1(1)+}^2X_{1(1)-}^2}}\; ,\\
\label{nilpotinv2}I_{2ij}&=&-\frac{\tilde{\bar{\Theta}}_{1(j)}\tilde{\mr X}_{1(2)-}\tilde{\Theta}_{1(i)}}{\sqrt{X_{1(1)+}^2X_{1(1)-}^2}}\; ,
}
which apart from $X_{1(1)+}^2$ and $X_{1(1)-}^2$ (eq. \eqref{X1ipm}) uses the following matrices and Gra\ss mannian spinors:
\eqa{
\label{X11plus}\tilde{\mr X}_{1(1)+}&=&-{\mr x_{\bar{1}4}}^{-1} \mr x_{\bar{2}4} {\mr x_{\bar{2}1}}^{-1}\; , \\
\label{X12minus}\tilde{\mr X}_{1(2)-}&=&{\mr x_{\bar{1}3}}^{-1} \mr x_{\bar{4}3} {\mr x_{\bar{4}1}}^{-1}\; , \\
\tilde{\Theta}_{1(i-1)}&=&\ci\rb{{\tilde{\mr x}_{\bar{4}1}}^{-1}\bar{\theta}_{41}-{\tilde{\mr x}_{\bar{i}1}}^{-1}\bar{\theta}_{i1}}\; ,\\
\tilde{\bar{\Theta}}_{1(i-1)}&=&\ci\rb{\theta_{1i}{\tilde{\mr x}_{\bar{1}i}}^{-1}-\theta_{14}{\tilde{\mr x}_{\bar{1}4}}^{-1}}\; .
}
Third powers of each of these spinors vanish, which restricts the number of possible monomials of the nilpotent invariants correspondingly. The variables introduced for the three point invariant in section \ref{sec3pI} are related to these variables by
\eqa{
\quad \quad \Theta_{1}&\!\!=\!\!&\Theta_{1(1)}-\Theta_{1(2)}\; ,\\
\bar{\Theta}_{1}&\!\!=\!\!&\bar{\Theta}_{1(1)}-\bar{\Theta}_{1(2)}\; ,\\
\tilde{\mr X}_{1+}&\!\!=\!\!&\tilde{\mr X}_{1(1)+}-\tilde{\mr X}_{1(2)-}-4\ci \bar{\Theta}_{1(2)}\Theta_{1(1)}\; .
}
It is convenient to divide eqns. \eqref{X11plus} and \eqref{X12minus} by a normalization similar to the one in the definition of the nilpotent invariants \eqref{nilpotinv1} and \eqref{nilpotinv2}:
\eqa{
\hat{\tilde{\mr X}}_{1(1)+}&=&\frac{\tilde{\mr X}_{1(1)+}}{\rb{X_{1(1)+}^2X_{1(1)-}^2}^{\frac{1}{4}}}\; ,\\
\hat{\tilde{\mr X}}_{1(2)-}&=&\frac{\tilde{\mr X}_{1(2)-}}{\rb{X_{1(1)+}^2X_{1(1)-}^2}^{\frac{1}{4}}}\; .
}
Again with the corresponding four-vectors one finds $J_{1(1)}$ in ${\hat{X}_{1(1)+}}^2$:
\equ{
{\hat{X}_{1(1)+}}^2=\sqrt{J_{1(1)}}=1-2\ci I_{111}+6 I_{111}^2\; .
}
Also $J_{1(2)}$, $\hat{X}_{1(1)+}^{\mu}\hat{X}_{1(2)-\mu}$ and $\hat{X}_{1(2)-}^2$ can be expressed in terms of the ten invariants \eqref{crorat}, \eqref{nilpotinv1} and \eqref{nilpotinv2}, which generate all four-point invariants. But here these rather long expressions are not needed. 

The following set $\mathfrak{I}$ of all linear independent monomials in the nilpotent invariants containing only 36 elements is chosen here:
\eqa{\label{nilpotinvset}
\mathfrak{I}_{0,1}&\in &\left\{ 1 \right\}\; ,\\ 
\nonumber \mathfrak{I}_{1,1\ldots 8}&\in &\left\{I_{ijk}\right\}\; ,\\ 
\nonumber \mathfrak{I}_{2,1\ldots 18}&\in & \{I_{1ij}I_{1kl}, I_{1ii}I_{212}, I_{1ii}I_{221}, \\
\nonumber && I_{111}I_{222}, I_{122}I_{211}, I_{i12}I_{221}\} \;,\\
\nonumber \mathfrak{I}_{3,1\ldots 8}&\in & \{ I_{111}^2 I_{i22}, I_{122}^2 I_{i11}, I_{112}^2 I_{j21},\\
\nonumber &&I_{121}^2 I_{j12}\} \; ,\\
\nonumber \mathfrak{I}_{4,1}&=& I_{111}^2 I_{122}^2
}
with $i, k,j, l=1,2$. They are sorted with the first index being the degree. 

This set is a subset of minimal size of the monomials of nilpotent invariants from eqns. \eqref{nilpotinv1} and \eqref{nilpotinv2}, so that any nilpotent invariant can be expressed as a linear combination of its elements with coefficients, which are functions of non-nilpotent invariants, e.g. the superconformal cross ratios. 

In the following the relations between the monomials, which have been left out in the choice of the subset, and the chosen ones are given using properties of the Pauli matrices and eqns. \eqref{ttequ1}-\eqref{ttequ2}. While both equations are applied to show 
\equ{
I_{1jk}I_{2jk}=\frac{\hat{X}_{1(1)+}^{\mu}\hat{X}_{1(2)-\mu}}{{\hat{X}_{1(1)+}}^2}I_{1jk}I_{1jk}
}
for $j,k=1,2$, only either one or the other can be used --  together with the symmetry of the contractions of $X_{1(2)-}^{\mu}$ with the Pauli-matrices hidden in the definition \eqref{nilpotinv2} -- to get 
\equ{
I_{2jk}I_{2mn}=\frac{{\hat{X}_{1(2)-}}^2}{{\hat{X}_{1(1)+}}^2}I_{1jk}I_{1mn}
}
for  $j,k,m,n=1,2$ with $j=m \vee k=n$.

$I_{211}I_{112}$, $I_{211}I_{121}$, $I_{112}I_{222}$ and $I_{121}I_{222}$ are equal to the expressions 
\eqa{&&  \\
\nonumber I_{211}I_{112}&=& 2 {\hat{X}_{1(1)+}}^{\mu} \hat{X}_{1(2)-\mu} I_{111}I_{112}-I_{111}I_{212}\; ,\\
\nonumber I_{211}I_{121}&=& 2 {\hat{X}_{1(1)+}}^{\mu} \hat{X}_{1(2)-\mu}  I_{111}I_{121}-I_{111}I_{221}\; ,\\
\nonumber I_{112}I_{222}&=& 2 {\hat{X}_{1(1)+}}^{\mu} \hat{X}_{1(2)-\mu} I_{112}I_{122}-I_{212}I_{122}\; ,\\
\nonumber I_{121}I_{222}&=& 2 {\hat{X}_{1(1)+}}^{\mu} \hat{X}_{1(2)-\mu}  I_{121}I_{122}-I_{221}I_{122}\; .
}
A bit longer are the relations giving $I_{121}I_{212}$,
\eqa{
&&I_{121}I_{212}=-I_{112}I_{221}-I_{111}I_{222}-I_{122}I_{211}\\
\nonumber&& +2 {\hat{X}_{1(1)+}}^{\mu} \hat{X}_{1(2)-\mu}\rb{I_{111}I_{122}+I_{112}I_{121}}\; ,
}
and $I_{211}I_{222}$,
\eqa{
&& I_{211}I_{222}=-I_{212}I_{221}\\
\nonumber&& +{\hat{X}_{1(2)-}}^2 \rb{I_{111}I_{122}+I_{112}I_{121}}\; .
}
In \cite{Park:1999pd} also nilpotent invariants with lowest order in Gra\ss mann variables $\rb{\theta\bar{\theta}}^2$ are constructed. This was necessary, as the  construction was made also for extended supersymmetry $\mathcal{N}>1$, which is not discussed here. For $\mathcal{N}=1$, for which eqns. \eqref{ttequ1}-\eqref{ttequ2} are valid, these invariants can be expressed by the nilpotent invariants above:   
\eqa{&& I_{1ij}^2\\
\nonumber &=&\frac{1}{4}{\hat{X}_{1(1)+}}^2 {\Theta}_{1(i)}\sigma^{\mu}\bar{\Theta}_{1(j)}{\Theta}_{1(i)}\sigma_{\mu}\bar{\Theta}_{1(j)}\; ,\\
&& I_{1ii}I_{112}\\
\nonumber &=&\frac{1}{4}{\hat{X}_{1(1)+}}^2 {\Theta}_{1(i)}\sigma^{\mu}\bar{\Theta}_{1(i)}{\Theta}_{1(1)}\sigma_{\mu}\bar{\Theta}_{1(2)}\; ,\\
 &&I_{1ii}I_{121}\\
\nonumber &=&\frac{1}{4}{\hat{X}_{1(1)+}}^2 {\Theta}_{1(i)}\sigma^{\mu}\bar{\Theta}_{1(i)}{\Theta}_{1(2)}\sigma_{\mu}\bar{\Theta}_{1(1)}\; ,\\
&& I_{111}I_{122} + I_{112}I_{121}\\
\nonumber &=&\frac{1}{2}{\hat{X}_{1(1)+}}^2 {\Theta}_{1(1)}\sigma^{\mu}\bar{\Theta}_{1(1)}{\Theta}_{1(2)}\sigma_{\mu}\bar{\Theta}_{1(2)}\; ,\\
\nonumber &=&\frac{1}{2}{\hat{X}_{1(1)+}}^2 {\Theta}_{1(1)}\sigma^{\mu}\bar{\Theta}_{1(2)}{\Theta}_{1(2)}\sigma_{\mu}\bar{\Theta}_{1(1)}
}
with $i,j=1,2$.

\section{Chiral superfields}
\label{sufi} Superfields are operator-valued distributions on superspace. They form representations of the superconformal group and have as quantum numbers the Lorentz spin, $(j_1,j_2)$, the scale dimension, $\eta$, and the R-charge, $\kappa$. Here only scalar chiral/anti-chiral superfields are discussed, i.e. $j_1=j_2=0$ and the restrictions
\equ{\label{restr1}
\bar{D}_+\Phi\rb{x_+^{\mu},\theta,\bar{\theta}}=0
}
for chiral fields, $\Phi$, and 
\equ{\label{restr2}
D_-\bar{\Phi}\rb{x_-^{\mu},\theta,\bar{\theta}}=0
}
for anti-chiral fields, $\bar{\Phi}$. Here chiral and anti-chiral coordinates and the corresponding covariant derivatives (eqns. \eqref{Dplus} and \eqref{Dminus}) are already used, which is very convenient, as these conditions then simply mean, that chiral and anti-chiral fields do not depend on $\bar{\theta}$ and $\theta$, respectively. They have the following expansions in $\theta$ and $\bar{\theta}$:
\eqa{
\quad \Phi\rb{{x}^{\mu}_{+},\theta}\!\!\!\!&=&\!\!\!\!\phi\rb{{x}^{\mu}_{+}}+\sqrt{2}\:\theta \tilde{\psi}\rb{{x}^{\mu}_{+}}\\
\nn &&+\theta\tilde{\theta}\:m\!\rb{{x}^{\mu}_{+}}\; ,\\
\quad \bar{\Phi}\rb{x_-^{\mu},\bar{\theta}}\!\!\!\!&=&\!\!\!\!\phi^*\rb{{x}_-^{\mu}}-\sqrt{2}\:\tilde{\bar{\theta}}\bar{\psi}\rb{{x}_-^{\mu}}\\
\nn &&-\tilde{\bar{\theta}}\bar{\theta}\:m^*\!\rb{{x}_-^{\mu}}\; .
}
From the superconformal transformations  of these superfields (cf. \cite{Park:1999pd}) it follows, that scale dimensions and R-charges of these fields are related as
\equ{\label{kappaeta}
\kappa=\pm \frac{1}{3}\eta \; ,
}
where $+$ and $-$ are valid for anti-chiral and chiral superfields, respectively. 

\section{Correlation functions}
\label{corr} Only the consequences of the superconformal symmetry and the restrictions of the superfields are discussed here without any reference to a specific model. Non-vanishing correlation functions of the component fields always have a vanishing total R-charge, which is the sum of the R-charges of the fields therein. So the total R-charge of a correlation function of superfields has to vanish for it to be non-nilpotent. For non-vanishing total R-charge and thus nilpotent correlation functions of scalar chiral superfields, the Ward identities seem to be rather restrictive: It was shown in \cite{Osborn:1998qu}, that the total R-charge of the three point function has to be $1$.

Here the nilpotent case will not be discussed. Therefore we have for the R-charges $\kappa_i$ of the superfields in a $n$-point function
\equ{\label{Rsum}
\sum_{i=1}^n \kappa_i=0 \; .
}
In general the scalar $n$-point function in superconformal field theory is a function of all invariants of $n$ points times a factor due to the superconformal transformations of the $n$ fields (cf. \cite{Park:1999pd}):
\equ{\label{npcorr}
\ang{S_1\ldots S_n}=\frac{F\rb{\text{n-point invariants}}}{\prod_{l,m=1;\;l \neq m}^{n}{x_{\bar{l}m}^2}^{\Delta_{lm}}}\; ,
}
where
\eqa{\label{Dlm}
&&\Delta_{lm}=-\frac{1}{2(n-1)(n-2)}
\sum_{i=1}^{n}\eta_{i}\\
\nonumber &&+\frac{1}{2(n-2)}(\eta_{l}+\eta_{m})+
\frac{3}{2n}
(\kappa_{l}-\kappa_{m})\; .
}

Here we are interested in chiral scalar superfields. As these depend only on the chiral variables, one immediately sees, that there is no way to construct a three point invariant with only half the Gra\ss mann variables. Therefore the three point function can be easily written down in the next section, as it was already done in e.g. \cite{Osborn:1998qu}. 

A four point function with vanishing total R-charge depends on two chiral and two anti-chiral variables (here $(x_{1-},\bar{\theta}_1)$, $(x_{2-},\bar{\theta}_2)$, $(x_{3+},\bar{\theta}_3)$ and $(x_{4+},\bar{\theta}_4)$), as will be seen in section \ref{4pf}. This leads to the problem, that there is only one superconformal cross ratio given by eq. \eqref{croratgen} depending only on these four variables, namely $\mathcal{I}_2$. But there have to be two independent non-nilpotent superconformal four-point invariants, because there are two conformal four-point invariants, on which the four-point functions of the component fields depend. One possibility is the construction of a trace invariant as the second non-nilpotent invariant (cf. \cite{Osborn:1998qu}).

Here two superconformal cross ratios, $\mathcal{I}_1$ and $\mathcal{I}_2$, and the set of nilpotent invariants $\mathfrak{I}$ from eq. \eqref{nilpotinvset} will be used to get an expression for the four-point functions. This will simplify conclusions from properties of global conformal field theories to global superconformal field theories. 

With the help of the chirality conditions applied to the four-point function, the dependence on all these invariants can be reduced to a dependence only on two cross ratios with a fixed universal differential operator applied to the whole rest of the correlation function. This rest contains the model specific information.

For rational four-point functions this differential operator can be applied to their power series. The coefficients of this power series turn out to be the same as those of the conformal four-point function of the scalar fields, which are the lowest components of the chiral and anti-chiral fields. 

\subsection{Two and three-point functions}
\label{23pf} As there are no two-point invariants, the two-point function is solely determined by the superconformal transformations properties of the two fields and a constant $C$:
\equ{
\ang{ \bar{\Phi}_1\rb{x_{1-}^{\mu},\bar{\theta}_1} 
\Phi_2\rb{x_{2+}^{\mu},\theta_2}} = C\frac{1}{(x_{\bar{1}2}^2)^{\eta}}\;.
}
From \eqref{xibj} one easily verifies the conditions on the two-point function given by the restrictions \eqref{restr1} and \eqref{restr2}, 
\eqa{
\ang{ D_{1-}\bar{\Phi}_1 \Phi_2}=0  && \ang{ \bar{\Phi}_1 \bar{D}_{2+}\Phi_2}=0\; .
}
The three-point function is also determined up to a constant just like the ordinary conformal three-point function:
\eqa{
\nonumber\ang{ \bar{\Phi}_1\rb{x_{1-}^{\mu},\bar{\theta}_1} \Phi_2\rb{x_{2+}^{\mu},\theta_2} \Phi_3\rb{x_{3+}^{\mu},\theta_3}}&& \\
 =\frac{C}{x_{\bar{1}2}^{2\eta_2}x_{\bar{1}3}^{2\eta_3}}\; .&&
}
Note, that $\eta_1=\eta_2+\eta_3$ for these three point functions because of eqns. \eqref{kappaeta} and \eqref{Rsum}.

As the strategy for the four-point function will be to look at all the invariants and then implement the chirality conditions, the following short calculation shows, how the chiral three point function fits into the form of the most general three point function (cf. eq. \eqref{npcorr}),
\eqa{
\nonumber \ang{ \bar{\Phi}_1\rb{x_{1-}^{\mu},\bar{\theta}_1} \Phi_2\rb{x_{2+}^{\mu},\theta_2} \Phi_3\rb{x_{3+}^{\mu},\theta_3}}&& \\
 =\frac{f_{3PF}\rb{I_{3PF}}}{\prod_{l,m=1;\;l \neq m}^{3}{x_{\bar{l}m}^2}^{\Delta_{lm}}}\; ,&&
}
which contains an arbitrary function of the three-point invariant, $I_{3PF}$.
The restrictions \eqref{restr1} and \eqref{restr2} and so the differential equations, 
\eqa{
\quad \quad D_{1-} \ang{\bar{\Phi}_1 \Phi_2 \Phi_3} = 0 \; ,\!\! && \!\! \bar{D}_{2+} \ang{ \bar{\Phi}_1 \Phi_2 \Phi_3} = 0\; , \\ \bar{D}_{3+} \ang{ \bar{\Phi}_1 \Phi_2 \Phi_3} = 0 \!\!\; , &&
}
have to be satisfied. The solution gets obvious, when the denominator is rewritten with the help of the identities relating the $\Delta_{lm}$ with each other:
\eqa{\frac{1}{\prod_{l,m=1;\;l \neq m}^{3}{x_{\bar{l}m}^2}^{\Delta_{lm}}} && \\
\nonumber = \frac{{J_1}^{\Delta_{21}}}{x_{\bar{1}2}^{2\eta_2}x_{\bar{1}3}^{2\eta_3}}\; .&&
}
The derivatives only vanish, if the whole does not depend on $I_{3PF}$ any more. So with eq. \eqref{I3PFJ1} the function $f_{3PF}\rb{I_{3PF}}$ is
\equ{
f_{3PF}\rb{I_{3PF}}=\rb{I_{3PF}+1}^{-\Delta_{21}}\; .
}

\subsection{Four-point functions}
\label{4pf}
First of all the general form of the scalar four-point function given by eq. \eqref{npcorr} for $n=4$ can be rewritten with the help of $\mathcal{I}_1$, $\mathcal{I}_2$, $J_1$, $J_{1(1)}$ and $J_{1(2)}$:
\eqa{
&&\!\!\ang{S_1\ldots S_4}=
\\ \nn&&{\mathcal{I}_1}^{\Lambda_{34}} {\mathcal{I}_2}^{\Lambda_{24}} {J_1}^{\Delta_{43}-\Delta_{31}}{J_{1(1)}}^{-\Delta_{43}-\Delta_{41}}\\
\nn && {J_{1(2)}}^{\Delta_{43}}\rb{\frac{{x_{\bar{1}2}}^2}{{x_{\bar{2}1}}^2}}^{\Sigma_1}\rb{\frac{{x_{\bar{2}3}}^2}{{x_{\bar{3}2}}^2}}^{\Sigma_2}\rb{\frac{{x_{\bar{2}4}}^2}{{x_{\bar{4}2}}^2}}^{\Sigma_3}\\
\nn && \rb{\frac{{x_{\bar{2}3}}^2}{{x_{\bar{1}2}}^2 {x_{\bar{1}3}}^2}}^{\Xi} \frac{F\rb{\mathcal{I}_1,\: \mathcal{I}_2,\: I_{111},\ldots,I_{222}}}{{x_{\bar{1}3}}^{2\rb{\eta_3 - \eta_2}}{x_{\bar{1}4}}^{2\eta_4}{x_{\bar{2}3}}^{2\eta_2}}
}
with 
\eqa{
 \quad \quad \Lambda_{lm}&=&\Delta_{lm}+\Delta_{ml}\\
 \nn &=&-\frac{1}{6}\sum_{i=1}^4 \eta_i  + \frac{1}{2} \rb{\eta_l+\eta_m}\; ,\\
 \label{defSig1}\Sigma_1&=&\Delta_{21}+\Delta_{31}+\Delta_{41}=\frac{1}{2}\eta_1-\frac{3}{2}\kappa_1\; ,\\
\label{defSig2}\Sigma_2&=&\Delta_{31}+\Delta_{32}+\Delta_{34}=\frac{1}{2}\eta_3-\frac{3}{2}\kappa_3\; ,\\
\label{defSig3}\Sigma_3&=&\Delta_{41}+\Delta_{42}+\Delta_{43}=\frac{1}{2}\eta_4-\frac{3}{2}\kappa_4\; ,\\
 \label{defXi}\Xi&=&\frac{1}{2}\rb{\eta_1+\eta_2-\eta_3-\eta_4}\; ,
 }
where eq. \eqref{Rsum} is used to calculate $\Sigma_1$, $\Sigma_2$ and $\Sigma_3$. Because the invariant prefactors pulled out can be expressed in terms of the invariants, which are the arguments of $F$, we can define a new function $f_{4PF}$, so that
\eqa{\label{Sca4PFgen}&&\\
\nn &&\ang{S_1\ldots S_4}\!\!=\!\!\rb{\frac{{x_{\bar{1}2}}^2}{{x_{\bar{2}1}}^2}}^{\Sigma_1}\rb{\frac{{x_{\bar{2}3}}^2}{{x_{\bar{3}2}}^2}}^{\Sigma_2}\rb{\frac{{x_{\bar{2}4}}^2}{{x_{\bar{4}2}}^2}}^{\Sigma_3}\\
\nn && \rb{\frac{{x_{\bar{2}3}}^2}{{x_{\bar{1}2}}^2 {x_{\bar{1}3}}^2}}^{\Xi}\frac{f_{4PF}\rb{\mathcal{I}_1,\: \mathcal{I}_2,\: I_{111},\ldots,I_{222}}}{{x_{\bar{1}3}}^{2\rb{\eta_3 - \eta_2}}{x_{\bar{1}4}}^{2\eta_4}{x_{\bar{2}3}}^{2\eta_2}}\; .
}
For four-point functions of anti-chiral and chiral superfields the four exponents $\Sigma_1$, $\Sigma_2$, $\Sigma_3$ and $\Xi$ vanish, as is seen below leading to eq. \eqref{4pfchi}.

In general the function $f_{4PF}$ has the expansion
\equ{\label{f4PFexpan}
f_{4PF}= \sum_{i=0}^{4} \sum_{j=1}^{n_i} f_{i,j}\rb{\mathcal{I}_1,\: \mathcal{I}_2} \mathfrak{I}_{i,j}
}
with respect to the nilpotent invariants \eqref{nilpotinvset}, where the coefficients are functions of the two cross ratios from eq. \eqref{crorat}. 

If the Gra\ss mann variables are set to zero, the superconformal four-point function passes into the conformal four-point function. The cross ratios, $\mathcal{I}_1$ and $\mathcal{I}_2$, reduce to conformal cross ratios and all $\mathfrak{I}_{i,j}$ vanish. Thus the function $f_{0,1}$ is the function appearing in the scalar four-point function of conformal field theory. 

In order to see, what consequences the restrictive conditions defining chiral and anti-chiral superfields have on the scalar four-point functions, eq. \eqref{Sca4PFgen}, and to get a general expression of the chiral/anti-chiral scalar four-point function with vanishing total R-charge, the following four differential equations have to be solved:
\eqa{
\label{Diff4pf}D_{1-}\ang{\bar{\Phi}\bar{\Phi}\Phi\Phi}&=&0\; ,\\
\nonumber D_{2-}\ang{\bar{\Phi}\bar{\Phi}\Phi\Phi}&=&0\; ,\\
\nonumber \bar{D}_{3+}\ang{\bar{\Phi}\bar{\Phi}\Phi\Phi}&=&0\; ,\\
\nonumber \bar{D}_{4+}\ang{\bar{\Phi}\bar{\Phi}\Phi\Phi}&=&0\; .
}
A four-point function with only one anti-chiral and three chiral fields (or vice versa) with vanishing total R-charge can be excluded with the corresponding differential equations already with lowest order calculations. 

Using $z_{i-}=\rb{x_{i-}^{\mu},\bar{\theta}_i}$ and $z_{i+}=\rb{x_{i+}^{\mu},\theta_i}$, the following theorem states the solution of the differential equations.
\begin{thm}
All chiral/anti-chiral scalar four-point functions with vanishing total R-charge are of the form
 \eqa{ \label{4pfres}
 & \ang{ \bar{\Phi}_1\rb{z_{1-}} \bar{\Phi}_2\rb{z_{2-}} \Phi_3\rb{z_{3+}} \Phi_4\rb{z_{4+}}}& \\
\nonumber &=e^{\mathcal{D}}\stb{\frac{f_{0,1}\rb{{\mathcal{I}_1}, {\mathcal{I}_2}}}{{x_{\bar{1}3}}^{2\rb{\eta_3 - \eta_2}}{x_{\bar{1}4}}^{2\eta_4}{x_{\bar{2}3}}^{2\eta_2}}}  &
  }
with the zeroth order coefficient, $f_{0,1}$, of the expansion of $f_{4PF}$ with respect to the nilpotent invariants (eq. \eqref{f4PFexpan}) and a fixed, universal differential operator 
\equ{\label{defD} \mathcal{D} = 4\ci\; T\rb{\mathcal{I}_1,\: \mathcal{I}_2,\: I_{111},\ldots,I_{222}} \; \mathcal{I}_1\der{\mathcal{I}_1}\; .
}
\end{thm}
The fixed function $T$ is nilpotent and will be given in the course of the proof in eq. \eqref{defT}. The exponential with the derivatives has to be understood in form of its Taylor expansion with only finitely many terms because of the nilpotency of $T$. 

\paragraph{\emph{Sketch of the proof:}} 
The starting point is the general scalar four point function. The chiral/anti-chiral scalar four-point function has to be of the same form, eq. \eqref{Sca4PFgen}. 

However, scale dimensions and R-charges of the chiral and anti-chiral superfields fulfill the eqns. \eqref{kappaeta} and \eqref{Rsum}. Consequently $\Sigma_1=\Sigma_2=\Sigma_3=\Xi=0$ (eqns. \eqref{defSig1}-\eqref{defXi}). Thus the four-point function reduces to
\equ{\label{4pfchi}
\ang{ \bar{\Phi}_1 \bar{\Phi}_2 \Phi_3 \Phi_4}=\frac{f_{4PF}\rb{\mathcal{I}_1,\: \mathcal{I}_2,\: I_{111},\ldots,I_{222}}}{{x_{\bar{1}3}}^{2\rb{\eta_3 - \eta_2}}{x_{\bar{1}4}}^{2\eta_4}{x_{\bar{2}3}}^{2\eta_2}}
}

The derivatives in the differential equations \eqref{Diff4pf} applied to the denominator of this form are zero. Thus for the differential equations to be satisfied the derivatives of the function $f_{4PF}$ have to vanish. These derivatives can be evaluated using the expansion \eqref{f4PFexpan}:
\eqa{\label{Diffeqexpan}
\mathfrak{D}f_{4PF} &=& \sum_{i=0}^{3} \sum_{j=1}^{n_i} \frac{\partial f_{i,j}}{\partial \mathcal{I}_1} \rb{\mathfrak{D}\mathcal{I}_1} \mathfrak{I}_{i,j} \\
\nn &&+ \sum_{k=1}^{4} \sum_{l=1}^{n_k} f_{k,l} \mathfrak{D} \mathfrak{I}_{k,l}
}
with $\mathfrak{D} \in \left\{D_{1-},D_{2-},\bar{D}_{3+},\bar{D}_{4+}\right\}$. The sum over $i$ ends at $i=3$ because the product of $\mathfrak{I}_{4,1}$ and the derivative of $\mathcal{I}_1$ is zero due to their nilpotency.

This suggests to look at the lowest a priori non-vanishing order, $\theta$ resp. $\bar{\theta}$, because it clarifies the relation between the function $f_{0,1}$ and the functions $f_{1,l}$: In this lowest order the term
\equ{\label{Diffeq1stord}
\frac{\partial f_{0,1}}{\partial \mathcal{I}_1} \rb{\mathfrak{D}\mathcal{I}_1}+\sum_{l=1}^{8} f_{1,l} \mathfrak{D} \mathfrak{I}_{1,l}
} 
has to vanish. 

So the problem breaks down to the cancelation of the derivatives of the four-point invariants in lowest order and one finds, that all $f_{1,l}$ are fully determined by $f_{0,1}$. 

To this order the differential equation is satisfied by
\eqa{\label{1stordsol}
\quad f_{4PF}&\!\!\!=\!\!\!& \rb{1 + 4 \ci T' \mathcal{I}_1 \der{\mathcal{I}_1}} f_{0,1}\rb{{\mathcal{I}_1},{\mathcal{I}_2}}\\
\nonumber &\!\!\!\!& +\mathcal{O}\rb{\rb{\theta\bar{\theta}}^2}\; ,
}
where 
\equ{\label{Tprime} T'=\frac{{\mathcal{I}_4}}{\sqrt{J_{1(1)}}}I_{222}+\frac{1}{\mathcal{I}_5\sqrt{J_{1(1)}}}I_{\Sigma}-\frac{\mathcal{I}_6}{\sqrt{J_{1(1)}}}I_{212}
}
is a fixed function of four-point invariants. The factors with cross ratios (eqns. \eqref{croratfur1}-\eqref{croratfur4}) and $J_{1(1)}$ (eq.\eqref{J112}) just eliminate overall prefactors of the nilpotent invariants, which simplifies calculations in lowest order. $I_{\Sigma}$ is a combination of all eight nilpotent invariants
\eqa{
\quad I_{\Sigma}\!\!\!&=\!\!\!&\frac{\tilde{\Theta}_{1}\tilde{\mr X}_{1+}\tilde{\bar{\Theta}}_{1}}{\sqrt{X_{1(1)+}^2X_{1(1)-}^2}}\\
\nonumber &=\!\!\!&-I_{111}+I_{112}+I_{121}-I_{122}\\
\nonumber &\!\!\!&+ I_{211}-I_{212}-I_{221}+I_{222}\\
\nonumber &\!\!\!&+4\ci ( 2 I_{112}^2 -2 I_{111}I_{112} \\
\nonumber &\!\!\!&-2 I_{112}I_{122} + I_{111}I_{122} + I_{112}I_{121})\\
 \nonumber &\!\!\!&-4 I_{111}^2 I_{222}-8 I_{112}^2 I_{121}\; ,
}
which is $I_{111}$ with the indices $4$ replaced by $3$ and is simply related to the three-point invariant in eq. \eqref{I3pAlt}.

Eq. \eqref{1stordsol} shows exactly, what restrictions follow from the lowest non-vanishing (i.e. first) order of the differential equations \eqref{Diff4pf}. If one looks beyond first order of Gra\ss mann variables, one simplifies matters even more going over to a new set of three nilpotent invariants replacing the three appearing in eq. \eqref{Tprime}:
\eqa{\label{T1t4}
\quad \quad \quad T_{222}\!\!\!\!&=\!\!\!\!&\frac{\mathcal{I}_4}{\sqrt{J_{1(1)}}}I_{222}-2\ci \frac{{\mathcal{I}_4}^2}{J_{1(1)}}I_{222}^2\\
\nonumber\!\!\!\!&=\!\!\!\!& -\rho_{14} + \rho_{13} +\rho_{34} \\
\nonumber \!\!\!\!&\!\!\!\!&+ 2 \ci \rb{{\rho_{14}}^2 - {\rho_{13}}^2 - {\rho_{34}}^2}\; ,\\
\quad \quad \quad T_{\Sigma}\!\!\!\!&=\!\!\!\!&\frac{I_{\Sigma}}{\mathcal{I}_5\sqrt{J_{1(1)}}}-2\ci \frac{I_{\Sigma}^2}{{\mathcal{I}_5}^2J_{1(1)}}\\
\nonumber\!\!\!\!&=\!\!\!\!& -\rho_{13} + \rho_{12} +\rho_{23} \\
\nonumber \!\!\!\!&\!\!\!\!&+ 2 \ci \rb{{\rho_{13}}^2 - {\rho_{12}}^2 - {\rho_{23}}^2}\; ,\\
\!\!\!\!&\!\!\!\!& \\
\nn T_{212}\!\!\!\!&=\!\!\!\!&\frac{\mathcal{I}_6}{\sqrt{J_{1(1)}}}I_{212}+2\ci \frac{\mathcal{I}_6^2}{J_{1(1)}}I_{212}^2\\
\nonumber\!\!\!\!&=\!\!\!\!& -\rho_{14} - \theta_{34}{\tilde{x}_{\bar{3}4}}^{-1}\bar{\theta}_{24} - \theta_{31}{\tilde{x}_{\bar{2}1}}^{-1}\bar{\theta}_{21} \\
\nonumber \!\!\!\!&\!\!\!\!&+ \theta_{34}{\tilde{x}_{\bar{3}4}}^{-1}{\tilde{x}_{\bar{2}3}}{\tilde{x}_{\bar{2}1}}^{-1}\bar{\theta}_{21}  \\
\nonumber \!\!\!\!&\!\!\!\!& + 2 \ci [\rho_{14}^2 + \rb{\theta_{34}{\tilde{x}_{\bar{3}4}}^{-1}\bar{\theta}_{24}}^2 \\
\nonumber \!\!\!\!&\!\!\!\!& - \!\! \rb{\theta_{31}{\tilde{x}_{\bar{2}1}}^{-1}\bar{\theta}_{21}}^2 \!\! + \!\! \rb{\theta_{34}{\tilde{x}_{\bar{3}4}}^{-1}{\tilde{x}_{\bar{2}3}}{\tilde{x}_{\bar{2}1}}^{-1}\bar{\theta}_{21}}^2\\
\nonumber \!\!\!\!&\!\!\!\!& - 2 \rb{\theta_{34}{\tilde{x}_{\bar{3}4}}^{-1}\bar{\theta}_{24} - \theta_{31}{\tilde{x}_{\bar{2}1}}^{-1}\bar{\theta}_{21} -\rho_{34}}\\
\nonumber \!\!\!\!&\!\!\!\!& \theta_{34}{\tilde{x}_{\bar{3}4}}^{-1}{\tilde{x}_{\bar{2}3}}{\tilde{x}_{\bar{2}1}}^{-1}\bar{\theta}_{21} - 2 \theta_{34}{\tilde{x}_{\bar{3}4}}^{-1}\bar{\theta}_{24} \rho_{34}]\\
\nonumber \!\!\!\!&\!\!\!\!& +16 \theta_{34}{\tilde{x}_{\bar{3}4}}^{-1}{\tilde{x}_{\bar{2}3}}{\tilde{x}_{\bar{2}1}}^{-1}\bar{\theta}_{21}\theta_{31}{\tilde{x}_{\bar{2}1}}^{-1}\bar{\theta}_{21}\\
\nonumber \!\!\!\!&\!\!\!\!& \rb{\rho_{34}- \theta_{34}{\tilde{x}_{\bar{3}4}}^{-1}\bar{\theta}_{24}}
}
with
\equ{
\label{rhoij} \rho_{ij} = \theta_{ij}{\tilde{x}_{\bar{i}j}}^{-1}\bar{\theta}_{ij}\; .
}
They have especially simple expressions in terms of the variables $\theta_i$, $\bar{\theta}_j$ and $\tilde{x}_{\bar{i}j}$ and differ  only in second and higher orders from those in eq. \eqref{Tprime}. 

As a side remark, in analogy $T_{111}$ can be defined, which even is exactly again $I_{111}$,
\eqa{
\quad \quad T_{111} \!\!\!&=\!\!\!&I_{111}\\
\nonumber\!\!\!&=\!\!\!& \frac{I_{111}}{\sqrt{J_{1(1)}}}-2\ci \frac{I_{111}^2}{J_{1(1)}}\\
\nonumber\!\!\!&=\!\!\!& -\rho_{14} + \rho_{12} +\rho_{24} \\
\nonumber \!\!\!&\!\!\!&+ 2 \ci \rb{{\rho_{14}}^2 - {\rho_{12}}^2 - {\rho_{24}}^2}\; .
}

Consequently $T'$ is replaced by
\equ{\label{defT}
T=T_{222}+T_{\Sigma}-T_{212}
}
which still solves the differential equations to first order. But now further terms in higher order vanish conveniently, because
\equ{\label{DIpropDT}
\mathfrak{D}\mathcal{I}_1=-4 \ci \mathcal{I}_1 \mathfrak{D} T\; .
}

The term in \eqref{Diffeq1stord} vanishes completely. Thus there are two summands less in eq. \eqref{Diffeqexpan}. The first order result and eq. \eqref{DIpropDT} can be plugged in, so that
\eqa{
&& \quad \quad \mathfrak{D}f_{4PF} = 16  \mathcal{I}_1 \der{\mathcal{I}_1}\rb{{\mathcal{I}_1} \frac{\partial f_{0,1}}{{\partial \mathcal{I}_1}}} T \mathfrak{D} T  \\
\nn && \!\!\!\!\!\!\!\!- \sum_{i=2}^{3} \sum_{j=1}^{n_i} \frac{\partial f_{i,j}}{\partial \mathcal{I}_1} \rb{4 \ci \:\mathcal{I}_1 \mathfrak{D} T} \mathfrak{I}_{i,j}
+ \sum_{k=2}^{4} \sum_{l=1}^{n_k} f_{k,l} \mathfrak{D} \mathfrak{I}_{k,l}
}
Note, that the coefficient functions $f_{k,l}$ with $k \geq 2$ have been redefined corresponding to the step, in which the nilpotent invariants were replaced. 

Looking at the lowest order left here ($\rb{\theta\bar{\theta}}\theta$ resp. $\rb{\theta\bar{\theta}}\bar{\theta}$) and using the product rule for $\mathfrak{I}_{2,l}$ it gets clear, that the same derivatives of four-point invariants need to cancel and thus the result can again only depend on $T$. 

This way it is possible to proceed iteratively order by order. The resulting function $f_{4PF}$ is
\eqa{
\quad f_{4PF}&\!\!\!=\!\!\!& \left[1 + 4 \ci T \mathcal{I}_1 \der{\mathcal{I}_1} \right.\\
\nn &&- 8 T^2 \rb{\mathcal{I}_1 \der{\mathcal{I}_1}\rb{\mathcal{I}_1 \der{\mathcal{I}_1}}} \\
\nn &&-\frac{32}{3}\ci \rb{\mathcal{I}_1 \der{\mathcal{I}_1}}^3\\
\nn &&\left.+ \frac{32}{3} \rb{\mathcal{I}_1 \der{\mathcal{I}_1}}^4\right] f_{0,1}\rb{{\mathcal{I}_1},{\mathcal{I}_2}} \\
\nn && +\mathcal{O}\rb{\rb{\theta\bar{\theta}}^2}\; .
}

As this is the expansion of the exponential in eq. \eqref{4pfres}, this finishes the proof. The dependence on nilpotent invariants has been totally fixed and the superconformal four-point function of scalar chiral and anti-chiral superfields is left with no more degrees of freedom than the scalar conformal four point function. 

The four point function in eq. \eqref{4pfres} is not explicitly given in terms of the chiral and anti-chiral coordinates $z_{1-}$, $z_{2-}$, $z_{3+}$ and $z_{4+}$. ${\mathcal{I}_1}$ and $\mathcal{D}$ depend on $\theta_2$ and $\bar{\theta}_3$. These terms, of cause, cancel.

\paragraph{\emph{Rational four-point functions:}} In the remaining part of this section the case of rational four-point functions is discussed, which will lead to an expression containing only $z_{1-}$, $z_{2-}$, $z_{3+}$ and $z_{4+}$. It was shown in \cite{Nikolov:2000pm}, that in global conformal field theory all correlation functions are rational. As a consequence of global conformal symmetry this is also the case for supersymmetric theories with this space-time symmetry. 

In rational four-point functions the function $f_{0,1}$ is a power series:
\equ{
f_{0,1}\rb{\mathcal{I}_1,\: \mathcal{I}_2} = \sum_{k_1,k_2}f_{0,1,k_1,k_2}{\mathcal{I}_1}^{k_1} {\mathcal{I}_2}^{k_2}\; . 
}
The differential operator applied to this leads to 
\equ{\label{DOonPS}
e^{\mathcal{D}} f_{0,1}\rb{\mathcal{I}_1,\: \mathcal{I}_2} = \sum_{k_1,k_2}f_{0,1,k_1,k_2} e^{4\ci\; T\; k_1}{\mathcal{I}_1}^{k_1} {\mathcal{I}_2}^{k_2}\; . 
}

This is rewritten now as a function of chiral and anti-chiral variables. Thus only $\theta_{34}$,  $\theta_{21}$ and $x_{\bar{i}j}$ with $i=1,2$ and $j=3,4$ may appear. Because of the exponential function in \eqref{DOonPS} it is useful to express $x_{\bar{1}2}$, $x_{\bar{3}4}$ and $x_{\bar{3}2}$ by the ''allowed`` intervals times an exponential function with nilpotent exponent. One finds, that
\eqa{
\label{relexp12} \quad \quad x_{\bar{1}2}^2\!\!\!\!&=\!\!\!\!&\rb{x_{\bar{1}3}^{\mu}-x_{\bar{2}3}^{\mu}}^{2}\\
\nonumber \!\!\!\!&\!\!\!\!& \cdot e^{4\ci\rb{\theta_{23}{\tilde{\mathrm x}_{\bar{1}2}}^{-1} \bar{\theta}_{21}-2\ci \rb{\theta_{23}{\tilde{\mathrm x}_{\bar{1}2}}^{-1} \bar{\theta}_{21}}^2}}\; ,\\
\label{relexp34} \quad \quad x_{\bar{3}4}^2\!\!\!\!&=\!\!\!\!&\rb{x_{\bar{1}3}^{\mu}-x_{\bar{1}4}^{\mu}}^{2}\\
\nonumber \!\!\!\!&\!\!\!\!& \cdot e^{4\ci\rb{\theta_{43}{\tilde{\mathrm x}_{\bar{3}4}}^{-1} \bar{\theta}_{13}-2\ci \rb{\theta_{43}{\tilde{\mathrm x}_{\bar{3}4}}^{-1} \bar{\theta}_{13}}^2}}\; ,\\
\label{relexplm} \quad \quad x_{\bar{3}2}^2\!\!\!\!&=\!\!\!\!& x_{\bar{2}3}^2 e^{-4\ci\rb{\rho_{23}-2\ci{\rho_{23}}^2}}
}
using $\rho_{23}$ from eq. \eqref{rhoij}. 

This is inserted into the cross ratio $\mathcal{I}_1$. The exponents in eqns. \eqref{relexp12}-\eqref{relexplm} multiplied with $k_1$ cancel most of the exponent in eq. \eqref{DOonPS}. The following four-point function remains:
\eqa{\label{4pfglconf}
 & \ang{ \bar{\Phi}_1\rb{z_{1-}} \bar{\Phi}_2\rb{z_{2-}} \Phi_3\rb{z_{3+}} \Phi_4\rb{z_{4+}}}& \\
\nonumber &=\frac{\sum_{k_1,k_2}f_{0,1,k_1,k_2} {\mathcal{I}_1'}^{k_1} {\mathcal{I}_2}^{k_2}}{{x_{\bar{1}3}}^{2\rb{\eta_3 - \eta_2}}{x_{\bar{1}4}}^{2\eta_4}{x_{\bar{2}3}}^{2\eta_2}} \;. &
}
with 
\equ{\label{I1prime}
\mathcal{I}_1' = e^{4 \ci \rb{\gamma - 2\ci \gamma^2}} \frac{x_{\bar{1}4}^2 x_{\bar{2}3}^2}{\rb{x_{\bar{1}3}^{\mu}-x_{\bar{2}3}^{\mu}}^{2} \rb{x_{\bar{1}3}^{\mu}-x_{\bar{1}4}^{\mu}}^{2}}
}
and
\eqa{\!\!\!&&\!\!\!\\
\nn \gamma\!\!\!&=&\!\!\!\frac{\theta_{34}\mathrm x_{\bar{1}4} \tilde{\mathrm x}_{\bar{1}3} \mathrm x_{\bar{2}3} \bar{\theta}_{21}}{\rb{x_{\bar{1}3}^{\mu}-x_{\bar{2}3}^{\mu}}^{2} \rb{x_{\bar{1}3}^{\mu}-x_{\bar{1}4}^{\mu}}^{2}} \\ 
\nn \!\!\!&-&\!\!\!\ \frac{{x_{\bar{1}3}}^2 \rb{\theta_{34} \mathrm x_{\bar{2}3} \bar{\theta}_{21} + \theta_{34}\mathrm x_{\bar{1}4} \bar{\theta}_{21} -  \theta_{34} \mathrm x_{\bar{1}3} \bar{\theta}_{21}}}{\rb{x_{\bar{1}3}^{\mu}-x_{\bar{2}3}^{\mu}}^{2} \rb{x_{\bar{1}3}^{\mu}-x_{\bar{1}4}^{\mu}}^{2}}\; .
}
Setting the Grassmann variables to zero, the exponent in \eqref{I1prime} vanishes and thus $I_1'$ passes into a conformal cross ratio just like $I_2$. Hence the coefficients $f_{0,1,k_1,k_2}$ are also those of the four-point function of the lowest component fields, which are scalar global conformal fields.

\section{Conclusion}
The proof of a general formula for the four-point function of chiral and anti-chiral superfields with vanishing total R-charge was sketched in this paper, while a long version will be available in my Ph.D.-thesis. In this formula the exponential of a universal nilpotent differential operator is applied to an arbitrary function of two superconformal cross ratios. This function will depend on the specific model. In the context of the proof a bunch of relations between superconformal invariants was presented. 

While research on the $\mathcal{N}=1$ superconformal four-point functions mainly focused on those of chiral or analytic superfields or on the analysis of component fields, the results here on the case of two anti-chiral and two chiral scalar superfields opens diverse  possibilities to transfer the knowledge on four-point functions of scalar conformal fields. As this four-point function has vanishing R-charge it is not nilpotent and reduces to the ordinary conformal case, if the Gra\ss mann variables are set to zero. Furthermore no degrees of freedom drop out in this step, so that there is a one-to-one correspondence between scalar conformal and these $\mathcal{N}=1$ superconformal four-point functions.

Especially interesting to transfer to superconformal theories are the results for global conformal invariance aiming at the rigorous construction of a non-trivial 4D quantum field theory in the setting of Wightman axioms. Both four-point functions of scalar chiral and anti-chiral superfields \emph{and} four-point functions of their lowest component fields have the same coefficients $f_{0,1,k_1,k_2}$ of their respective power series (cf. \eqref{4pfglconf}). Pole bounds \cite{Nikolov:2000pm} and an exact partial wave expansion computed for the scalar four-point function \cite{Nikolov:2005yu} are also valid for lowest component fields of chiral and anti-chiral superfields in a global superconformal theory, i.e. they give further restrictions on the coefficients $f_{0,1,k_1,k_2}$. Consequently these are results for global conformal just as for global superconformal four-point functions.

\subsection*{Remark:} In a very recent paper \cite{Poland:2010wg} the four-point function of two chiral and two anti-chiral superfields all having the same scaling dimension are given using a trace invariant. This trace invariant is given by
\equ{
\mr{tr}\rb{{\mr x_{\bar{2}3}}^{-1}{\mr x_{\bar{1}3}}{\mr x_{\bar{1}4}}^{-1}{\mr x_{\bar{2}4}}}=1-{\mathcal{I}_1'}^{-1}+{\mathcal{I}_2}^{-1}
}
with ${\mathcal{I}_1'}$ and ${\mathcal{I}_2}$ from eqns. \eqref{I1prime} and \eqref{crorat}, respectively. If this invariant is used and the rational four-point function is written as a power series as in eq. \eqref{4pfglconf}, the coefficients would be resorted in comparison to eq. \eqref{4pfglconf} and the step from global conformal to global superconformal four-point function would be not so direct.
%\appendix
%\input{appv2}
%\newpage
%\section{Muell}
%\input{muell}
\bibliographystyle{./hunsrt}
\bibliography{sc4pf}
\address{\onecolumn
Institut f\"ur Theoretische Physik, Georg-August-Universit\"at G\"ottingen \newline%
\indent Friedrich-Hund-Platz 1, D-37077 G\"ottingen, Germany}%
\email{Holger.Knuth@theorie.physik.uni-goettingen.de}%
\urladdr{http://www.theorie.physik.uni-goettingen.de/$\sim$knuth}
\end{document}